\documentclass[pdflatex,sn-nature]{sn-jnl}

\usepackage[T1]{fontenc}
\usepackage{lmodern}
\usepackage{graphicx}
\usepackage{amsmath,amssymb,amsfonts}
\usepackage{textcomp}
\usepackage{booktabs}
\usepackage{longtable,array}
\usepackage{siunitx}
\usepackage{placeins}
\usepackage{microtype}
\usepackage{bookmark}

\hypersetup{hidelinks}
\makeatletter
\renewcommand{\normalsize}{%
  \@setfontsize\normalsize{11}{14}%
  \abovedisplayskip=9pt plus 2pt minus 2pt
  \belowdisplayskip=9pt plus 2pt minus 2pt
  \abovedisplayshortskip=3pt plus 2pt
  \belowdisplayshortskip=6pt plus 2pt minus 2pt}
\makeatother
\normalsize

\setcitestyle{super,sort&compress}
\begin{document}

\title[Electron Wavefronts Capture Nanoscale Movies of Hypersound at Surfaces]{\protect\centering Electron Wavefronts Capture Nanoscale Movies of Hypersound at Surfaces}

\author[1,2,3]{\fnm{Johannes} \sur{Otto}}%

\author[1,2]{\fnm{Leon} \sur{Brauns}}%
\author[1,2]{\fnm{Benjamin} \sur{Schröder}}%
\author*[1,2,3]{\fnm{Claus} \sur{Ropers}}\email{claus.ropers@mpinat.mpg.de}

\affil[1]{\orgdiv{Ultrafast Dynamics}, \orgname{Max Planck Institute for Multidisciplinary Sciences}, \orgaddress{\street{Am Faßberg 11}, \city{Göttingen}, \postcode{37077}, \state{Lower Saxony}, \country{Germany}}}

\affil[2]{\orgdiv{IV. Physics Institute}, \orgname{Georg-August-Universität Göttingen}, \orgaddress{\street{Friedrich-Hund-Platz 1}, \city{Göttingen}, \postcode{37077}, \state{Lower Saxony}, \country{Germany}}}

\affil[3]{ \orgname{Max Planck School of Photonics}}

\maketitle

\noindent{\bfseries
Acoustic waves transport mechanical energy and elastic strain through solids, with material composition and shape governing their propagation. In the gigahertz hypersound regime\cite{maldovan2013}, acoustic wavelengths reach tens to hundreds of nanometres, comparable to nanoscale device dimensions. Harnessing these waves underpins high-frequency signal processing\cite{giribaldi2024}, sensing\cite{li2023,mandal2022a}, electromagnetic coupling\cite{mou2025,hwang2024a} and quantum control\cite{satzinger2018a,bienfait2019a}. Acoustic metrology probes elastic modes and internal interfaces\cite{ng2022,krenner2026,illienko2025a}. Yet directly imaging how nanoscale acoustic waves interact with local structure remains challenging, requiring picometre displacement sensitivity combined with high spatiotemporal resolution. Here we demonstrate full-field imaging of laser-induced hypersound at surfaces using reflected electron wavefronts. We introduce ultrafast low-energy electron microscopy (ULEEM), exploiting the intrinsic surface sensitivity of few-electronvolt electrons. Coherent electron pulses down to femtosecond durations reflect from the topmost atomic layers, imprinting the evolving surface profile onto their wavefronts. Defocus converts phase variations into image contrast. In a thin van der Waals material, we image the thermoelastic generation and propagation of picometre-amplitude acoustic waves. Measured dispersion relations, interpreted through elastodynamic modelling, distinguish substrate-supported and locally detached regions, revealing how interfacial contact shapes propagation. Delayed wave emission reveals scattering at buried mechanical defects. Our results establish electron-wavefront imaging of nanoscale surface dynamics, enabling the exploration of ultrafast energy transport, structural transformations and coupled mechanical, electronic and magnetic phenomena.}

\section{Introduction}

Surface-sensitive detection of acoustic waves provides access to structure and dynamics across an extraordinary range of length scales. Seismic waves reveal subsurface structures in geophysical exploration\cite{virieux2009}, while medical ultrasound forms images with submillimetre resolution from echoes generated at internal interfaces\cite{wells2006}. In solids, elastic waves transport mechanical energy and strain and increasingly serve as tools for probing and controlling materials properties\cite{maldovan2013}. At micro- and nanoscale dimensions, surface acoustic waves (SAWs) enable radio-frequency filtering\cite{giribaldi2024}, sensing\cite{mandal2022a}, and couple mechanical motion to electronic\cite{mou2025}, optical\cite{li2023,selvin2025}, magnetic\cite{hwang2024a} and quantum systems\cite{satzinger2018a,bienfait2019a}. Advances in both fundamental studies and applications depend on understanding acoustic-wave generation, propagation, scattering and dissipation in heterogeneous media\cite{krenner2026,ferrari2023}.

Resolving hypersound requires the combination of subatomic displacement sensitivity, nanometre spatial resolution and ultrafast timing\cite{xie2019,ng2022}. Ultrafast scanning-probe and near-field microscopies have reached atomic to nanometre length scales and femtosecond time resolution\cite{esses2026ultrafast}, including direct imaging of coherent molecular motion\cite{cocker2016tracking}. Scanning-probe techniques map stationary nanoscale acoustic fields\cite{hellemann2022,ng2022,lee2023}, while coherent X-ray and extreme-ultraviolet diffraction imaging reconstruct transient lattice and surface deformations\cite{clark2013,karl2018}. With even shorter wavelengths, electron microscopy offers complementary access through diffraction\cite{cremons2016,zhou2023,feist2018,lagrange2025} and secondary-electron contrast\cite{najafi2018}. Photoemission contrast has been used to map the piezoelectric potentials of standing SAWs\cite{foerster2019a,foerster2017}. Direct full-field imaging of propagating nanoscale surface waves remains challenging.

These advances support efforts to relate acoustic scattering and mode conversion to local structure and device performance, a technologically highly relevant direction. Direct full-field imaging of transient acoustic surface deformations with deep nanometre spatial resolution, picosecond temporal resolution and picometre displacement sensitivity, however, remains an outstanding challenge \cite{krenner2026}.

In this article, we introduce ultrafast low-energy electron microscopy (ULEEM) for directly imaging nanoscale acoustic surface dynamics. By converting a conventional low-energy electron microscope\cite{tromp2010a,bauer2014} to pulsed operation, we combine nanometre-scale imaging with the pronounced surface sensitivity of few-electronvolt electrons. We use this capability to visualize the surface response of a layered material following quasi-homogeneous ultrafast excitation. The measurements resolve transient surface displacements associated with the generation, propagation and scattering of elastic waves, revealing their dispersion relations and subsurface defects, and linking nanoscale acoustic dynamics to local geometry and substrate contact.

\section{Results}

\begin{figure}[tb]
    \centering
\includegraphics[width=.56\linewidth]{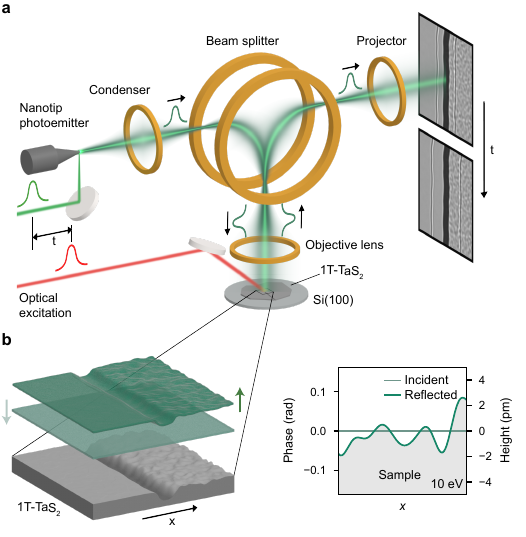}

\caption{\textbf{Ultrafast low-energy electron microscopy.}
\textbf{a,} Schematic of the ULEEM setup. A 1T-TaS$_2$ flake is heated by an ultrafast laser pulse and imaged by an electron pulse photoemitted from a nanotip at an adjustable delay $t$.
\textbf{b,} The surface height profile $h(\mathbf r)$ introduces a geometrical path difference and local phase shift $\phi(\mathbf r)$ in the reflected electron wave. Weakly defocusing the objective lens converts this phase modulation into image contrast.}
\label{fig:uleem-principle}
\end{figure}

A schematic of the experiment is shown in Fig.~\ref{fig:uleem-principle}a. The output of a laser system operating at \qty{610}{\kilo\hertz} repetition rate is split into ultrashort pump and probe pulses. After frequency doubling to \qty{515}{\nano\metre}, the probe pulses are focused onto a nanotip in a Schottky-type emitter assembly, generating ultrashort electron pulses by photoemission\cite{feist2017}. The electron pulses are accelerated to \qty{15}{\kilo\electronvolt}, directed towards the sample by a magnetic beam splitter, and collimated to provide plane-wave illumination at normal incidence over a diameter of approximately \qty{5}{\micro\metre}. A retarding potential between the objective lens and the sample decelerates the electrons to a landing energy $E_0$ of a few electronvolts, limiting their interaction with the sample to the topmost atomic layers\cite{bauer2014}. The reflected electrons pass through the objective lens a second time and are redirected by the beam splitter into a projection system, forming a magnified image on a microchannel-plate detector. Further details of the instrument design and performance are provided in Methods and Extended Data Fig.~\ref{fig:extended-data-uleem-performance}.

To resolve dynamic nanoscale surface displacements, we employ phase-contrast imaging of a structurally heterogeneous and elastically strongly anisotropic sample comprising exfoliated 1T-TaS$_2$ flakes on a Si(100) substrate. As illustrated in Fig.~\ref{fig:uleem-principle}b, the surface topography $h(\mathbf{r})$ introduces a spatially varying path length, which is directly imprinted onto the phase front of the reflected electron wave as \cite{bauer2014}
\begin{align*}
\phi(\mathbf{r}) = \frac{4\pi h(\mathbf{r})}{\lambda_0}.
\end{align*}
At a landing energy of \qty{10}{\electronvolt}, the de Broglie wavelength $\lambda_0$ is approximately \qty{3.88}{\angstrom}. Consequently, a small displacement of only \qty{1}{\angstrom} results in a phase shift of approximately $\pi$. As coherent imaging techniques can reveal much smaller phase shifts than this, electron reflection microscopy becomes exceptionally sensitive to sub-Å surface displacements.

In conventional electron and optical microscopy, phase variations of the wavefront are not directly visible in the intensity of an idealized (aberration-free) in-focus image. Changing the objective-lens excitation introduces an electron-optical defocus $\Delta z$, which converts local phase curvature into image-intensity variations $I(\mathbf r,\Delta z)$. For a surface with a homogeneous reflection coefficient, the transport-of-intensity equation\cite{teague1983a} yields
\begin{align}
  \label{eq:contrastTIE}
I(\mathbf r,\Delta z) &\propto \Delta z \,\frac{\lambda}{\lambda_0}\nabla_\perp^2h(\mathbf r).
\end{align}
Here, $\lambda\approx\qty{10.0}{\pico\metre}$ is the electron wavelength at the \qty{15}{\kilo\electronvolt} beam energy in the projection system. This relation provides an intuitive description of the resulting phase contrast in the small-defocus limit, with image contrast directly visualizing surface curvature. For quantitative image simulations, we additionally account for partial coherence and higher-order aberrations using an established Fourier-optics model\cite{schramm2012c,bauer2014}, as detailed in Methods.

\begin{figure}[tb]
\includegraphics[width=1\linewidth]{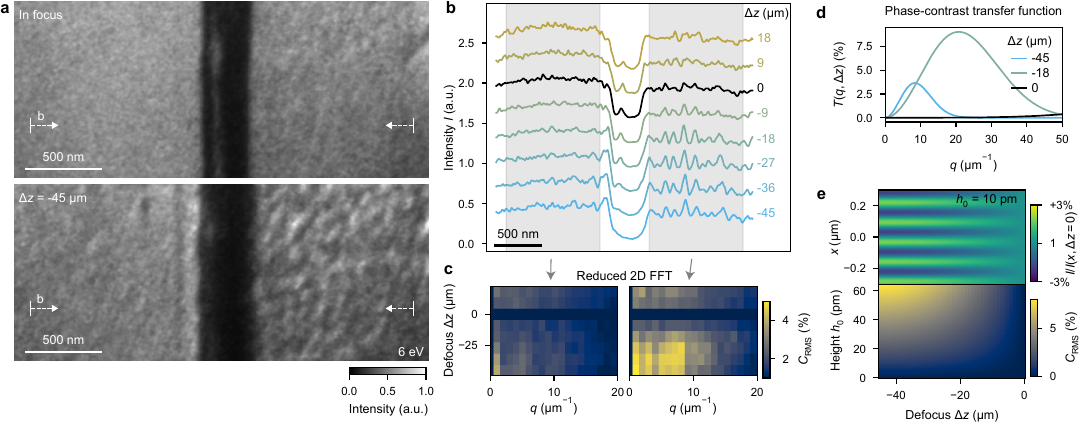}
\caption{\textbf{Defocus-dependent imaging of static surface topography.}
\textbf{a,} In-focus and defocused LEEM images of a crack in a 1T-TaS$_2$ flake. Dashed arrows mark the profile direction and the 100 nm transverse averaging width used in b.
\textbf{b,} Intensity profiles at different defocus values, vertically offset for clarity.
\textbf{c,} Fourier analysis of the regions on either side of the crack, using the convention defined in Methods.
\textbf{d,} Effective phase-contrast transfer function $T(q,\Delta z)$, including partial coherence, for different defocus settings. 
\textbf{e,} Forward-modelled image intensity for a sinusoidal surface ripple $h_0 \sin(2 \pi q_c x)$ with amplitude $h_0$ and spatial frequency $q_c=\qty{8.1}{\per\um}$. The upper map shows increasing visibility and saturation of the ripple contrast. The lower map gives the root-mean-square (RMS) contrast as a function of ripple amplitude and defocus.}
\label{fig:static-topography}
\end{figure}

Figure~\ref{fig:static-topography}a demonstrates how local height variations produce contrast in a defocus image series of the investigated van der Waals material\cite{pang2013a,yu2017a}. In the in-focus image, the regions on either side of a fracture in the 1T-TaS$_2$ flake show little discernible difference. Upon introducing defocus, pronounced periodic contrast appears on the right-hand side of the crack, revealing a static surface ripple. Its spatial periodicity of about \qty{100}{\nano\metre} to \qty{150}{\nano\metre} and the defocus dependence are evident in the Fourier transform of an intensity line profile across the structure (Fig.~\ref{fig:static-topography}c). At the dominant spatial frequency $q_m=\qty{8.1}{\per\um}$, the observed \qty{4.3}{\percent} contrast corresponds to a peak surface corrugation amplitude of approximately \qty{35}{\pico\metre} through using the modelled contrast-transfer function (Fig.~\ref{fig:static-topography}e), which demonstrates sensitivity to subatomic surface displacements. Importantly, a suitably chosen defocus can provide strong phase contrast down to wavelengths of few nanometres\cite{tromp2010a}, corresponding to the particularly relevant regime of high-frequency nanoacoustics and phononics\cite{krenner2026}.

\begin{figure}[tb]
\includegraphics[width=1\linewidth]{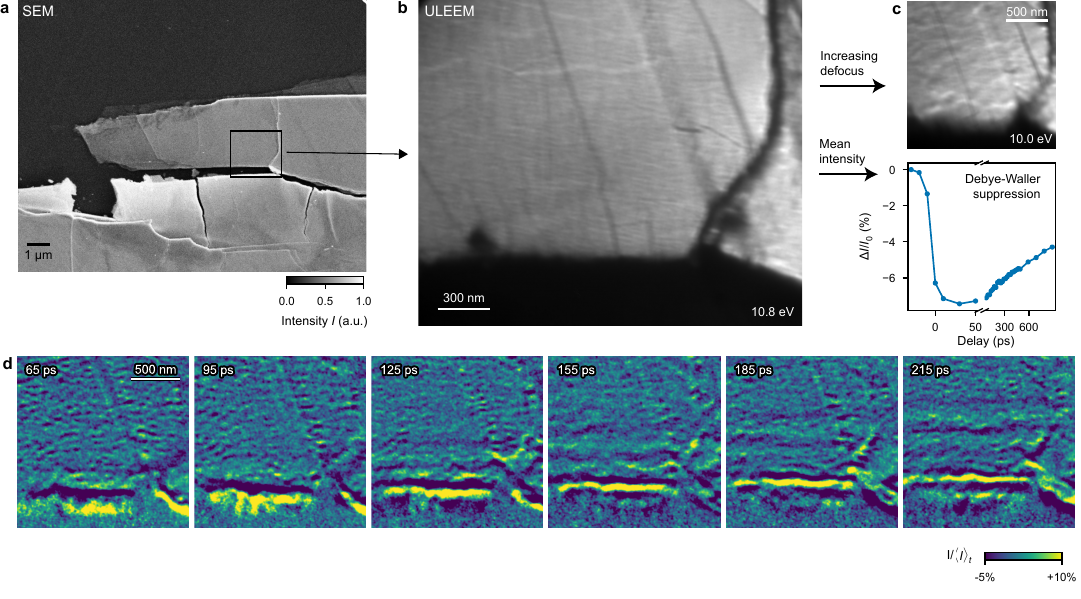}

\caption{\textbf{Ultrafast thermoelastic surface dynamics.}
\textbf{a,} Scanning electron micrograph of an exfoliated 1T-TaS$_2$ flake on a silicon substrate (black).
\textbf{b,} LEEM image of the TaS$_2$ flake acquired with the pulsed electron beam.
\textbf{c,} Defocused image (top) and spatially averaged relative intensity change $\Delta I/I_0$ after laser excitation (bottom). The intensity decreases transiently because of Debye-Waller suppression, and subsequently recovers.
\textbf{d,} Time-resolved image series after laser excitation, showing emerging wavefronts and smaller spatial intensity fluctuations.}
\label{fig:thermoelastic-dynamics}
\end{figure}

We now exploit the ultrafast temporal resolution of the instrument to directly map the generation and propagation of laser-induced ultrasonic waves. The first measurements focus on a \qty{70}{\nm}-thick region of a 1T-TaS$_2$ flake, independently characterized by scanning electron microscopy (Fig.~\ref{fig:thermoelastic-dynamics}a) and atomic force microscopy (Extended Data Fig.~\ref{fig:extended-data-model-validation}a). Panels b and c of Fig.~\ref{fig:thermoelastic-dynamics} display LEEM images acquired at increasing defocus, revealing a subtle long-range waviness of the surface. In the time-resolved experiments, the sample is excited quasi-homogeneously across the probed region with \qty{1030}{\nm} laser pulses (see Methods). ULEEM bright-field movies are recorded as a function of the delay between the optical pump and electron probe pulses.

After laser excitation, a transient global suppression and recovery of the image intensity is visible in Fig.~\ref{fig:thermoelastic-dynamics}c, stemming from laser-induced heating and the resulting Debye-Waller suppression of the surface reflection\cite{chase2016,kurtz2024}. We also find a sequence of travelling wavefronts with different group velocities launched from the sample edge and propagating across the surface (Fig.~\ref{fig:thermoelastic-dynamics}d). Smaller transient contrast ripples are visible across the imaged region. These images resolve acoustic wavefronts with wavelengths of 100 nm and below, substantially below the diffraction limit of visible-light microscopy, and track their propagation relative to local surface features.

\begin{figure}[tb]
\includegraphics[width=1\linewidth]{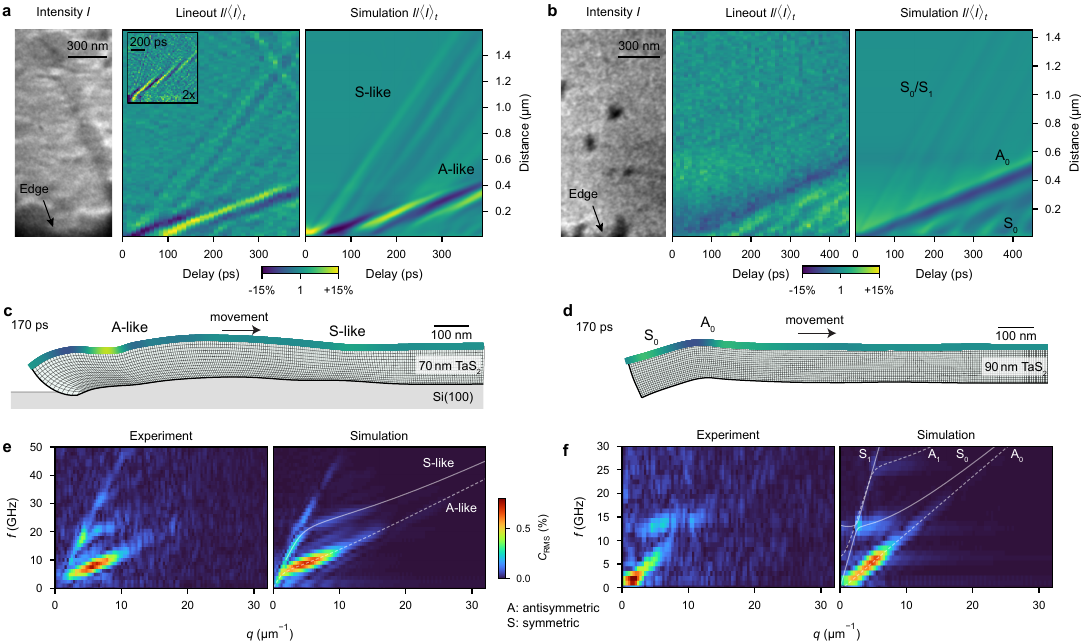}
\caption{\textbf{Imaging of edge-launched wave dynamics.}
\textbf{a,} Region of interest in a substrate-supported \qty{70}{\nano\metre}-thick flake (left). Experimental space-time map of waves propagating away from the flake edge (centre). $\langle I\rangle_{t>0}$ denotes the mean intensity at each pixel after laser excitation. The inset shows a long-delay measurement with the displayed contrast enhanced by a factor of 2. Simulated space-time map of the contrast for the heterostructure (right).
\textbf{b,} Region exhibiting a different propagation pattern, analysed with the same normalization and compared with a simulation of a free-standing \qty{90}{\nano\metre}-thick TaS$_2$ flake.
\textbf{c,d,} Simulated deformation of the TaS$_2$ flakes with (c) and without (d) a substrate, respectively, with displacements amplified by a factor of 100. The surface colour represents the simulated intensity $I/\langle I\rangle_{t>0}$, largely representing local surface curvature.
\textbf{e,f,} Two-dimensional Fourier transforms of the experimental and simulated space-time maps in a and b, revealing the wave dispersion. Lines show semi-analytical dispersion relations for elastic modes of different character in each geometry.}
\label{fig:edge-wave-model}
\end{figure}

To examine how these waves are generated, we record ULEEM movies at higher magnification near the flake edge (Fig.~\ref{fig:edge-wave-model}a, left).
We normalize the intensity at each point in the image by its temporal mean after excitation, $I/\langle I\rangle_{t>0}$. Averaging this normalized intensity along the edge yields the space-time map in Fig.~\ref{fig:edge-wave-model}a, centre. The contrast modulation reaches \qty{20}{\percent} and comprises repeatedly launched fast wavefronts ($f=\qty{18}{\giga\Hz}$, $v=\qty{4.7}{\km\per\s}$) together with a slower mode of higher amplitude at a frequency $f\approx\qty{8}{\giga\Hz}$.
The latter wave is strongly dispersive, as shown by phase fronts that propagate notably faster than the envelope (phase velocity $v_{ph}=\qty{1.4}{\km\per\s}$ compared with a group velocity $v_{gr}=\qty{1.2}{\km\per\s}$), see also the zoomed-out inset.

These propagating wavefronts can be traced to the elastic response of the specimen following laser excitation\cite{matsuda2015a}. The abrupt temperature increase generates substantial thermoelastic stress throughout the specimen. At surfaces, boundaries and defects, the local stress imbalance leads to the generation of strain waves. Their propagation is set by the elastic stiffness tensor and mass density, while the flake thickness and substrate contact shape their dispersion\cite{nakamura2023,greener2019,karl2018}.

Generally, a flake of finite thickness on a substrate is expected to support guided modes with Rayleigh- and Lamb-like character\cite{karl2018,xie2019}, whose dispersion relations can be calculated for continuous films and multilayers\cite{arminhuber2026}. For our specific geometry and excitation conditions, we simulate the response to rapid heating using a two-dimensional finite-element model that solves the coupled elastodynamic and transient heat-conduction equations (Methods). We use theoretically determined  elastic constants for 1T-TaS$_2$\cite{bastos2019}. For direct comparison with the measurements, the calculated surface-normal displacement is passed through the numerical Fourier-optics model to predict LEEM intensity at each time delay. Because the simulated intensity depends approximately linearly on fluence and defocus, we adjust simulation parameters within reasonable bounds to match the observed intensity, starting from experimental calibrations (see Supplemental Information). The simulated time evolution of the ULEEM contrast for the AFM-measured film thickness is shown in Fig.~\ref{fig:edge-wave-model}a, right. The model closely reproduces the experimentally observed fast and slow waves as well as their relative amplitudes and many other subtle features.

A snapshot of the simulated lattice strain and surface curvature at a delay of \qty{170}{\pico\second} after excitation illustrates the types of modes propagating across the film (Fig.~\ref{fig:edge-wave-model}c). In particular, the fast wavefront displays largely symmetric strain with respect to the midplane of the film, while the slower feature shows a strain profile that increases from the bottom of the film to the surface.

The close correspondence of experiment and modelling is further supported by comparing the respective dispersion relations (Fig.~\ref{fig:edge-wave-model}e), obtained by spatiotemporal Fourier transforms. Both measurement and simulation exhibit signatures of several branches. An overlay with semi-analytical dispersion relations for an infinite TaS$_2$ film on silicon allows us to assign the fast and slow waves to fundamental Rayleigh-Lamb modes with primarily symmetric (S) and antisymmetric (A) character, respectively, partially modified by the presence of the substrate. Notably, the dispersive character of the A-like wave, evident from an upward frequency shift of this branch, directly stems from the substrate contact.

We observe such ultrasonic waves in multiple regions on the specimen, with variations in local amplitude and specific wave patterns, illustrating the method's sensitivity to structural heterogeneity and differences in substrate contact. In the measurement shown in Fig.~\ref{fig:edge-wave-model}b, centre, the response differs markedly from that discussed above. A longer-lived oscillation is observed near the sample edge, whereas the fast wave is not resolved. Most strikingly, the pulse leading the slower wavefront propagates without dispersion at nearly the same group velocity as the slow mode in Fig.~\ref{fig:edge-wave-model}a, centre. This behaviour suggests that the flake is locally decoupled from the substrate near the edge. To test this interpretation, we retain the same literature elastic tensor and model the region as a free-standing plate. Close agreement with the observed wave pattern is obtained for a free-standing \qty{90}{\nm}-thick TaS$_2$ plate (Fig.~\ref{fig:edge-wave-model}b, right), with the same defocus as before. The simulated response contains several symmetric and antisymmetric Lamb modes (Fig.~\ref{fig:edge-wave-model}d). Comparison of the experimental and simulated dispersion relations (Fig.~\ref{fig:edge-wave-model}f) identifies the dominant propagating feature as the fundamental antisymmetric A$_0$ mode, which is nearly nondispersive over the experimentally accessed wavevector range.

This assignment therefore provides evidence that the flake has little or no local substrate contact near this edge. The nondispersive character of the mode is directly visible in the experimental dispersion map (Fig.~\ref{fig:edge-wave-model}f, left), where its spectral weight follows a linear branch passing through the origin. The simulated dispersion map (Fig.~\ref{fig:edge-wave-model}f, right) further reveals a nearly flat spectral branch associated with localized standing waves near the flake edge. These breathing-like modes have negligible group velocity and correspond to the free-standing flake undergoing repeated out-of-plane oscillations at a frequency close to \qty{13}{\giga\hertz}.

\begin{figure}[tb]

\includegraphics[width=1\linewidth]{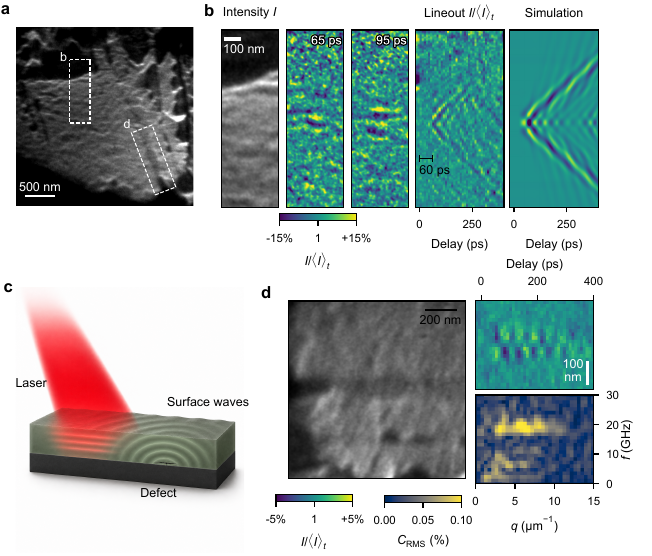}

\caption{\textbf{Delayed elastic responses at buried inhomogeneities.}
\textbf{a,} Static LEEM overview of the regions imaged in b and d. 
\textbf{b,} Intensity image with little spatial variation before laser excitation (left). Pixel-wise intensity changes at selected pump-probe delays (centre). Experimental space-time map compared with a simulation of a 30 nm-wide region at the underside of the flake (right), revealing a delayed onset of traveling surface modulations.
\textbf{c,} Schematic of an initially uniform out-of-plane strain pulse that scatters at a buried structural inhomogeneity and returns to the surface with a lateral modulation. 
\textbf{d,} Static view of a buried structure, with an intensity space-time map and its Fourier transform revealing a delayed oscillatory response.}

\label{fig:buried-dynamics}

\end{figure}

Beyond the waves launched directly at the flake edge, we observe both propagating and standing-wave responses in regions that appear rather homogeneous in the static surface image (Fig.~\ref{fig:buried-dynamics}a). Notably, the associated contrast emerges only after approximately \qty{60}{\ps}, with propagating components frequently radiating outward from otherwise featureless locations on the specimen (Fig.~\ref{fig:buried-dynamics}b). We associate this type of response with subsurface scattering of an initially uniform out-of-plane strain pulse. Pulsed optical excitation launches strong compressional strain pulses normal to a free surface\cite{matsuda2015a,illienko2025a}. Because ULEEM contrast is sensitive to lateral variations in surface curvature, the initially uniform strain pulse remains largely invisible and becomes detectable only after scattering introduces a lateral modulation and the pulse returns to the surface (see schematic in Fig.~\ref{fig:buried-dynamics}c).

Using the literature stiffness $C_{33}$ and mass density $\rho$, we calculate the longitudinal sound velocity $v_c=\sqrt{C_{33}/\rho}\approx\qty{2.38}{\km\per\s}$ along the crystallographic $c$ axis. Considering the approximate \qty{60}{\ps} round-trip time, we obtain a depth of  $\approx\qty{70}{\nm}$, in agreement with the AFM-measured thickness of the flake. Regions of incomplete surface contact, for example at buried steps or defects, are consistent with mechanical exfoliation, which can introduce cracks and other structural inhomogeneities at either surface of the flake.

A distinct manifestation of the sensitivity to subsurface structure is provided by a standing-wave response confined to a narrow, elongated region of the film (Fig.~\ref{fig:buried-dynamics}d). This region forms a shallow, barely discernible depression, while atomic force microscopy confirms that the surface remains continuous and intact (Extended Data Fig.~\ref{fig:extended-data-model-validation}a). Following optical excitation, a strongly localized oscillatory contrast emerges after a similar delay, with a characteristic frequency of about \qty{20}{\GHz}. Its strong spatial confinement and nearly flat dispersion identify it as a localized standing mode. Together with its elongated geometry and delayed onset, this response strongly indicates a buried exfoliation step that locally lifts the flake from the substrate, allowing the unsupported region to oscillate as a nanoscale membrane.

\FloatBarrier

\section{Discussion}

The demonstrated ability to image picometre-scale surface motion with nanometre spatial and picosecond temporal resolution provides direct access to nanoscale elastic-wave dynamics. By resolving wave propagation in both substrate-supported and locally free-standing regions, ULEEM reveals how finite thickness, interfacial contact and nanoscale boundaries shape acoustic dispersion, propagation and scattering. Delayed wave emission and localized resonances reveal otherwise hidden mechanical inhomogeneities, with their spatial origins and arrival times offering lateral localization and sonar-like depth ranging.

Beyond the contrast-based analysis presented here, through-focus image series combined with iterative phase retrieval\cite{yu2010a,duden2014a} can provide a fully quantitative reconstruction of the transient two-dimensional surface topography. Moreover, aberration-corrected photoemission-based LEEM \cite{wan2017b} could in the future enable ultrafast surface imaging down to 2 nm spatial resolution. ULEEM thereby offers a route to understanding how nanoscale heterogeneity governs energy flow, confinement and dissipation in layered materials, membranes and nanoscale devices. Other LEEM contrast mechanisms could extend these studies to phase transitions, magnetization dynamics and interactions with localized electromagnetic fields, providing complementary views of how mechanical deformation couples to electronic and magnetic order.

\bibliography{sn-bibliography}
\FloatBarrier

\clearpage
\section*{Methods}
\subsection*{Ultrafast low-energy electron microscopy}

The instrument is based on a commercial low-energy electron microscope modified for pulsed operation. A custom nanotip-based photoelectron gun is coupled into the microscope at the position of the original electron source through an additional magnetic prism. The source consists of a Schottky-type W(100) emitter operated below the threshold for continuous electron emission. Upon laser illumination, photoelectrons are generated and accelerated to a beam energy of \qty{15}{\kilo\electronvolt}. The additional prism couples the electron pulses into the conventional electron-optical beam path, with an intermediate crossover between the two prisms providing an approximately achromatic transfer to the sample. In the cathode-lens field between the objective lens and sample, the electrons are decelerated to a landing energy $E_0$ of a few electronvolts and illuminate an area of approximately \qty{5}{\micro\metre} in diameter. After reflection from the sample, the electrons pass through the objective lens a second time and are redirected into the projection system, where they are detected using a microchannel plate, phosphor screen and CMOS camera. The modified instrument retains the conventional imaging, spectroscopy and diffraction modes of the LEEM, including energy-filtered and dark-field imaging as well as micro-LEED. Measurements are performed under ultrahigh-vacuum conditions at a pressure of about \qty{5e-10}{\milli\bar}. Representative instrument parameters and performance are shown in Extended Data Fig.~\ref{fig:extended-data-uleem-performance}.

The photoelectron source is driven by \qty{515}{\nano\metre} laser pulses obtained by frequency doubling the output of a \qty{1030}{\nano\metre} Yb regenerative amplifier operating at a repetition rate of \qty{610}{\kilo\hertz}. The laser output is split into pump and probe arms with independently controlled intensities. The fundamental \qty{1030}{\nano\metre} pump pulses are p-polarized and focused onto the sample at an angle of incidence of \qty{70}{\degree}, yielding a calculated short-axis \(1/e^2\) intensity diameter of approximately \qty{45}{\micro\metre}. The relative arrival time of the optical pump and electron probe pulses is varied using a motorized optical delay stage in the probe arm. The pump focus on the sample and probe focus on the nanotip are actively stabilized. Depending on the experimental conditions, the optical pulse duration is adjusted with the internal laser compressor to values between \qty{0.2}{\pico\second} and \qty{7.8}{\pico\second}. Detailed excitation conditions and electron-optical settings for individual measurements are provided in the Supplementary Information.

\subsection*{Sample preparation and characterization}

Using adhesive tape, we mechanically exfoliated bulk 1T-TaS$_2$ crystals (HQ Graphene) \textit{ex situ} onto Si(100) substrates with native oxide. Before exfoliation, the substrates were rinsed with isopropanol and dried under N$_2$. After tape removal, the samples were transferred immediately to the load-lock and annealed at approximately \qty{450}{\kelvin} until the pressure stabilized in the ultrahigh-vacuum regime.

After the ULEEM measurements, the samples were characterized by scanning electron microscopy and atomic force microscopy (AFM). The micrograph in Fig.~\ref{fig:thermoelastic-dynamics}a was acquired with a secondary-electron detector at an accelerating voltage of \qty{5.00}{\kilo\volt}. The AFM height map in Extended Data Fig.~\ref{fig:extended-data-model-validation}a was used to determine the local flake thickness.

\subsection*{Forward model of image formation}

Image formation in the microscope is simulated using a one-dimensional Fourier-optics formalism\cite{altman1998,schramm2012c}. We use the contrast transfer function
\begin{align*}
H(q,\Delta z,\epsilon)&=M(q)\exp\!\left[i2\pi\chi(q,\Delta z,\epsilon)\right], \qquad M(q)=\begin{cases}1, & |q|\leq q_M,\\0, & |q|>q_M,\end{cases},
\end{align*}
where $M(q)$ is the aperture function with cutoff $q_M$. The aberration phase is given by
\begin{align*}
\chi(q,\Delta z,\epsilon)&=\frac{1}{4}\left[-2\left(\Delta z-C_{\mathrm{c}}\frac{\epsilon}{E}\right)\lambda q^2+C_3\lambda^3q^4\right],
\end{align*}
where $\Delta z$ is the defocus, $C_{\mathrm{c}}$ is the chromatic aberration coefficient and $C_3$ is the spherical aberration coefficient. Here, $E$ and $\lambda$ denote the electron energy and wavelength in the column.

Partial coherence is described by averaging over the illumination angle $\alpha$ relative to the optical axis and the energy offset $\epsilon$ relative to the central beam energy. Both follow normalized Gaussian distributions $s(u,\sigma)=1/(\sqrt{2\pi}\sigma)\exp[-u^2/(2\sigma^2)]$,
where $\sigma$ is the standard deviation. The detector intensity is
\begin{align*}
I(x,\Delta z)&=\int\mathrm{d}\alpha\,s(\alpha,\sigma_{\alpha})\int\mathrm{d}\epsilon\,s(\epsilon,\sigma_{\epsilon})
\left|\mathcal{F}^{-1}\!\left\{H\!\left(q+\frac{\alpha}{\lambda},\Delta z,\epsilon\right)\hat{\psi}(q,\epsilon)\right\}\right|^2.
\end{align*}
Here, $\hat{\psi}(q,\epsilon)$ denotes the Fourier transform of the object wave $\psi(x,\epsilon)$ at spatial frequency $q$. The defocus calibration, aberration coefficients and coherence parameters used in the simulations are given in the Supplementary Information.

The spatially varying surface height $h(x)$ introduces the phase shift
$\phi(x,\epsilon)=4\pi h(x)/\lambda_0(\epsilon)$,
where $\lambda_0(\epsilon)$ is the electron wavelength at the landing energy. Assuming a constant reflection amplitude $A_0$, the object wave is $\psi(x,\epsilon)=A_0\exp\!\left[i\phi(x,\epsilon)\right]$.

Under the weak-phase approximation, $\exp(i\phi)\approx1+i\phi$, and neglecting higher-order terms in the angular and energy spreads, we introduce the temporal and spatial coherence envelopes\cite{schramm2012c}
\begin{align*}
E_{\mathrm{C}}(q)&=\exp\!\left[-\frac{1}{2}\left(\pi C_{\mathrm{c}}\lambda q^2\frac{\sigma_{\epsilon}}{E}\right)^2\right],\\
E_{\mathrm{S}}(q,\Delta z)&=\exp\!\left[-2\pi^2\sigma_{\mathrm{\alpha}}^2\left(C_3\lambda^2q^3-\Delta z\,q\right)^2\right].
\end{align*}
The approximate image intensity is then
\begin{align*}
I(x,\Delta z) &\approx A_0^2 \Bigl\{1-2\mathcal{F}^{-1}\Bigl[
\underbrace{M(q)E_{\mathrm{C}}(q)E_{\mathrm{S}}(q,\Delta z)\sin\!\bigl(2\pi\chi(q,\Delta z,0)\bigr)}_{T(q,\Delta z)}
\hat{\phi}(q)\Bigr]\Bigr\}.
\end{align*}
The effective contrast transfer function $T(q,\Delta z)$ is plotted in Fig.~\ref{fig:static-topography}d. For Fig.~\ref{fig:static-topography}e, we used the full numerical forward model to calculate the image intensity for a sinusoidal height modulation $h(x)=h_0\sin(2\pi q_c x)$. The spatial frequency $q_c=\qty{8.1}{\per\um}$ matches the dominant component in the defocus scan. The simulated profiles were analysed with the same Fourier procedure as the experimental data, as described below.

\subsection*{Data acquisition and analysis}

For each data set, multiple image series were acquired, registered\cite{jong2020} and averaged. The mean detector background was determined from an unexposed region of the microchannel plate and subtracted from each image. Unless stated otherwise, the pump-probe movies were corrected for the Debye-Waller effect by dividing each frame by its mean intensity.
For a pair of coordinates $(a,b)$ and their conjugate frequencies $(q_a,q_b)$, the two-dimensional Fourier transform of the contrast $u(a,b)$ is defined as
\begin{align}
\tilde{u}(q_a,q_b)
&=
\sum_{a,b}
u(a,b)
\exp\left[-2\pi i(q_a a+q_b b)\right].
\end{align}
With $N_a$ and $N_b$ denoting the numbers of samples along each coordinate, the root-mean-square (RMS) Fourier amplitude is
\begin{align}
A_{\mathrm{RMS}}(q_a,q_b)
&=
\sqrt{2}
\left|
\frac{\tilde{u}(q_a,q_b)}
{N_aN_b}
\right|.
\end{align}

For the Fourier analysis of the defocus series, we selected regions on either side of the crack. Each spanned the full image height and the corresponding grey area marked in the line profiles in Fig.~\ref{fig:static-topography}b. The images in these regions were divided pixel-wise by the in-focus image and Fourier transformed in two dimensions. Taking the RMS along the direction perpendicular to the ripples yielded the one-dimensional Fourier spectrum in Fig.~\ref{fig:static-topography}c. For the simulated ripple profiles described above, the RMS contrast was evaluated within the spatial-frequency interval $q_c\pm\qty{1}{\per\um}$ and plotted in Fig.~\ref{fig:static-topography}e.

Time-resolved image series were normalized by the mean intensity at each pixel after laser excitation. Averaging the normalized intensity parallel to the edge over the indicated region of interest yielded the space-time maps. The wave dispersion was obtained from a zero-padded two-dimensional Fourier transform of these maps. To reduce artefacts, we applied a Kaiser window with $\beta=3$ along the temporal axis and compensated for the resulting amplitude reduction by multiplying the Fourier amplitudes by approximately 1.5.

\subsection*{Elastodynamic simulations}
The generation and propagation of laser-induced elastic waves are modeled by solving the coupled transient heat-conduction and linear elastodynamic equations in COMSOL Multiphysics 6.3 using a two-dimensional plane-strain finite-element model. The displacement field $\mathbf{u}$ evolves according to
\begin{align*}
\rho\frac{\partial^2\mathbf{u}}{\partial t^2}
&=\nabla\cdot\boldsymbol{\sigma}
\quad \text{with}\quad
\boldsymbol{\sigma}=\mathbf{C}\left[\boldsymbol{\varepsilon}-\boldsymbol{\varepsilon}_{\mathrm{th}}\right],
\end{align*}
where $\rho$ is the mass density, $\boldsymbol{\sigma}$ is the stress tensor, and $\mathbf{C}$ is the elastic stiffness tensor. The strain $\boldsymbol{\varepsilon}$ is derived from the displacement field $\mathbf{u}$, while $\boldsymbol{\varepsilon}_{\mathrm{th}}$ denotes the thermal eigenstrain.
The thermal eigenstrain is $\boldsymbol{\varepsilon}_{\mathrm{th}}=\boldsymbol{\alpha}(T-T_0)$, where $\boldsymbol{\alpha}$ is the thermal-expansion tensor and $T_0$ is the initial temperature. The transient temperature $T$ follows the heat-conduction equation,
\begin{align*}
\rho C_p\frac{\partial T}{\partial t}
&=\nabla\cdot\left(\mathbf{k}\nabla T\right)+Q,
\end{align*}
where $C_p$ is the specific heat capacity, $\mathbf{k}$ is the thermal-conductivity tensor, and $Q$ is the deposited power density.

The optical excitation is represented by a volumetric heat source in the TaS$_2$ flake with an exponential absorption profile normal to the surface and a Gaussian temporal envelope~\cite{thomsen1986a},
\begin{align*}
Q(z,t)
&=\frac{(1-R)F}{\xi\sigma_t\sqrt{2\pi}}
\exp\!\left[-\frac{(t-t_0)^2}{2\sigma_t^2}\right]
\exp\!\left(-\frac{z}{\xi}\right),
\end{align*}
where $F$ is the incident laser fluence, $R$ is the reflectivity, $\xi$ is the optical absorption length, and $z$ is the depth below the surface. The Gaussian envelope has a standard deviation $\sigma_t$ and is centred at $t_0$.

1T-TaS$_2$ is treated as elastically and thermally anisotropic, with the crystallographic $c$ axis normal to the flake surface, whereas the Si substrate is treated as isotropic. We consider substrate-supported and free-standing geometries corresponding to the experimentally investigated regions, with the flake thickness set to the independently measured local thickness. Structural inertia is retained to describe wave propagation, whereas geometric nonlinearities, mechanical losses, and thermoelastic damping are neglected. 

The thermoelastic simulations use a surface mesh spacing of \qty{2}{\nano\metre}. At each time point, displacements are added to the mesh coordinates before the surface is interpolated onto a uniform \qty{2}{\nano\metre} in-plane grid. The resulting height profiles are passed to the forward model to calculate the ULEEM contrast.

Dispersion relations were calculated using Dispersion Calculator v3.2\cite{arminhuber2026} with geometries and material parameters similar to those in the COMSOL simulations. The solver combines analytical elastic-wave solutions through transfer- and stiffness-matrix methods. It numerically determines modes that satisfy continuity of displacement and traction at interfaces and traction-free conditions at outer surfaces.

Material parameters and their sources, optical excitation conditions, simulation geometries and numerical settings are listed in the Supplementary Information.

\subsection*{Statistics and reproducibility}

The measurements in Figs.~2-5 were acquired from different regions of one 1T-TaS$_2$ flake on a single specimen. Time-resolved images were formed by stroboscopic accumulation over repeated pump-probe cycles at each delay. Propagating elastic waves were observed in both the substrate-supported and locally decoupled regions compared in Fig.~\ref{fig:edge-wave-model}.

\subsection*{Use of artificial intelligence tools}

We used large language models and associated coding tools to assist with code development, exploratory data analysis, and language editing. The authors reviewed and verified all AI-assisted code and analyses and retain full responsibility for the scientific results and conclusions.

\bmhead{Data availability}

The data supporting the findings of this study are available from the corresponding author upon request.

\backmatter

\bmhead{Acknowledgements}

We acknowledge insightful conceptual discussions with R. Tromp, M. Horn-von Hoegen, F.-J. Meyer zu Heringdorf, J. G. Horstmann, R. Haindl, A. Feist, H. Böckmann-Clemens, M. El-Gomati, J. Jobst, K. Kunze, and O. Schaff. We thank P. Schädlich, T. Seyller, T. de Jong and S. J. van der Molen for helpful advice on LEEM operation and sample preparation. We also thank T. Diedrichs, O. Böttger, F. Long, N. Bach, F. Kurtz, S. Briesenick and other members of the Ultrafast Dynamics department for technical support and useful discussions.

\bmhead{Funding}

We are grateful to the Max Planck School of Photonics for financial support. This project was funded by the European Research Council (ERC) under the European Union's Horizon Europe research and innovation programme (ERC Advanced Grant, Grant Agreement No.~101055435, ULEEM).

\bmhead{Author contributions}

C.R. and J.O. conceived the study and designed the experiments. All authors contributed to the development of the ULEEM instrument. L.B. and B.S. prepared the samples. J.O. and L.B. performed the measurements and analysed the data. J.O. developed the analysis software and curated the data and code. J.O. and C.R. developed the theory, performed the simulations and interpreted the results. J.O. prepared the figures and wrote the first draft. J.O. and C.R. revised the manuscript. C.R. supervised the research and secured funding.

\bmhead{Competing interests}

C.R. is a founder and shareholder of QSEM GmbH, a startup company developing ultrafast electron-beam instrumentation. J.O., L.B. and B.S. hold indirect financial interests in the company through inventor remuneration. C.R. and J.O. are co-inventors on the published international patent application WO~2024/188471~A1 concerning ultrafast low-energy electron microscopy.

\bmhead{Additional information}

Supplementary Information is available for this paper.

Correspondence and requests for materials should be addressed to Claus Ropers (\href{mailto:claus.ropers@mpinat.mpg.de}{claus.ropers@mpinat.mpg.de}).

\clearpage

\onecolumn

\section*{Extended Data}

\renewcommand{\theHfigure}{ED\arabic{figure}}
\setcounter{figure}{0}
\renewcommand{\figurename}{Extended Data Fig.}
\begin{figure}[!ht]
\centering
\includegraphics[width=\linewidth]{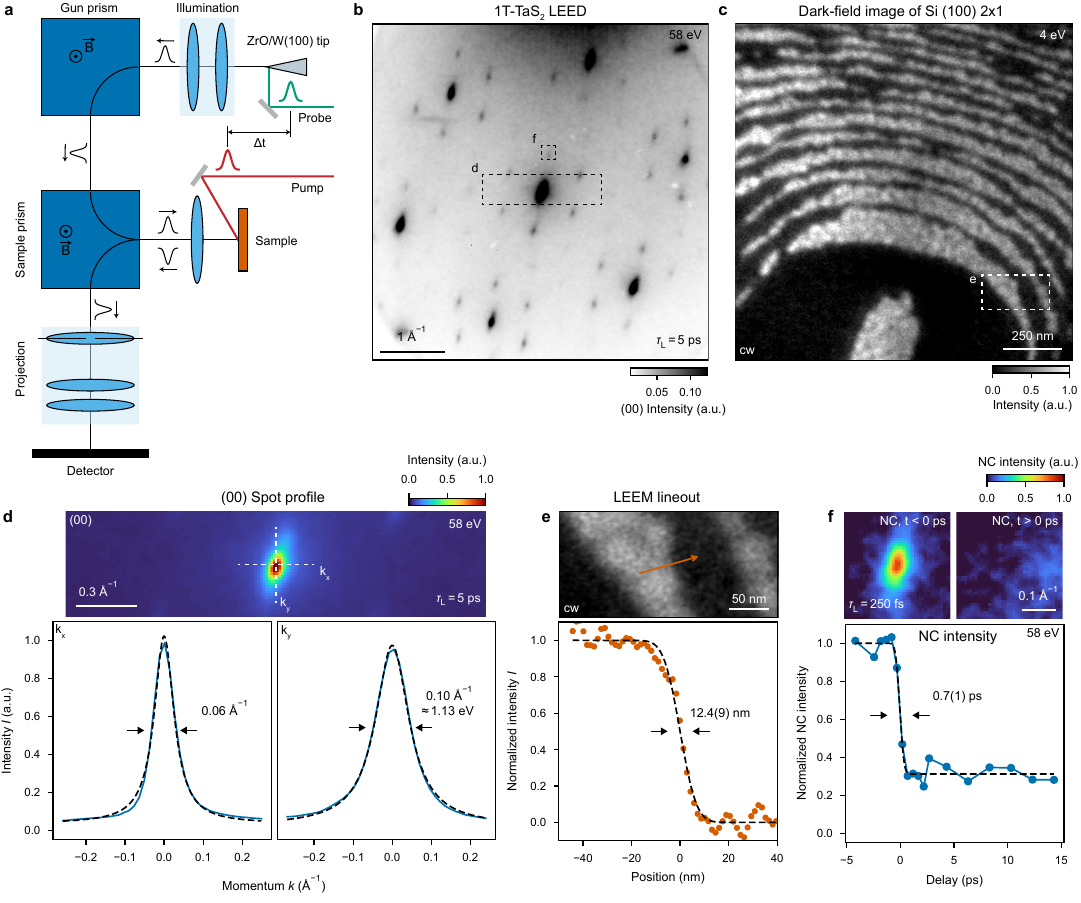}
\caption{\textbf{ULEEM setup and instrument performance.}
\textbf{a,} Detailed Electron-optical layout.
\textbf{b,} Electron diffraction pattern recorded using photoemission laser pulses with $\tau_L = \qty{5}{\pico\second}$ and normalized to the (00) diffraction intensity.
\textbf{c,} Continuous-wave photoemission dark-field LEEM image of the $2\times1$ reconstruction of Si(100)\cite{bauer2014}, used to benchmark the spatial resolution.
\textbf{d,} Selected diffraction spot and line profiles along $k_x$ and $k_y$. Dashed curves show Lorentzian fits with full widths at half maximum (FWHM) of \qty{0.06}{\per\angstrom} and \qty{0.10}{\per\angstrom}, respectively. The width along $k_y$ corresponds to an estimated energy width of \qty{1.13}{\electronvolt}.
\textbf{e,} Line profile from the dark-field image. An error-function fit yields a spatial resolution of \qty{12.4(9)}{\nano\metre} FWHM.
\textbf{f,} Ultrafast diffraction measurement of the transition from the nearly commensurate to the incommensurate phase\cite{Vogelgesang2018} in 1T-TaS$_2$ using laser pulses with $\tau_L = \qty{250}{\femto\second}$. The measurement demonstrates an instrument response below \qty{0.7(1)}{\pico\second} FWHM.}
\label{fig:extended-data-uleem-performance}
\end{figure}
\clearpage
\begin{figure}[!ht]
\centering
\includegraphics[width=\linewidth]{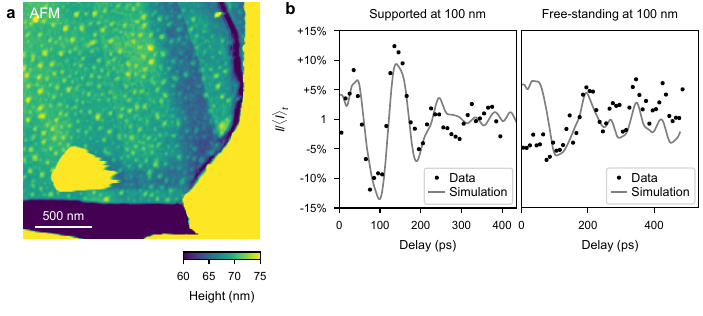}
\caption{\textbf{AFM characterization and analysis of the modelled regions.}
\textbf{a,} Atomic force microscopy height map used to determine the local flake thickness in the supported region. \textbf{b,} Experimental (black points) and simulated (grey curves) normalized intensity $I/\langle I\rangle_{t>0}$ at a distance of \qty{100}{\nano\metre} from the edge. The left and right panels show the supported and locally decoupled regions, respectively, with the same pixel-wise temporal normalization as in Fig.~\ref{fig:edge-wave-model}a,b. Percentage labels indicate deviations from unity.}
\label{fig:extended-data-model-validation}
\end{figure}

\end{document}


\renewcommand{\thepage}{S\arabic{page}}

\title[Supplementary Information]{Supplementary Information for\\Electron Wavefronts Capture Nanoscale Movies of Hypersound at Surfaces}

\author[1,2,3]{\fnm{Johannes} \sur{Otto}}
\author[1,2]{\fnm{Leon} \sur{Brauns}}
\author[1,2]{\fnm{Benjamin} \sur{Schröder}}
\author*[1,2,3]{\fnm{Claus} \sur{Ropers}}\email{claus.ropers@mpinat.mpg.de}

\affil[1]{\orgdiv{Ultrafast Dynamics}, \orgname{Max Planck Institute for Multidisciplinary Sciences}, \orgaddress{\street{Am Faßberg 11}, \city{Göttingen}, \postcode{37077}, \state{Lower Saxony}, \country{Germany}}}
\affil[2]{\orgdiv{IV. Physics Institute}, \orgname{Georg-August-Universität Göttingen}, \orgaddress{\street{Friedrich-Hund-Platz 1}, \city{Göttingen}, \postcode{37077}, \state{Lower Saxony}, \country{Germany}}}
\affil[3]{\orgname{Max Planck School of Photonics}}

\maketitle

\section*{Supplementary Methods}

\subsection{Measurement settings}

Table~\ref{tab:measurement-settings} lists the measurement settings and associated figures for each data set. All measurements used a \qty{20}{\micro\metre} contrast aperture. The optical pulse duration $\tau_L$ is the intensity full-width-at-half-maximum (FWHM). For a Gaussian beam, the estimated peak incident fluence $F$ at the sample surface is
\begin{align*}
    F = \frac{2E_{\mathrm{pulse}}}{A},
\end{align*}
where $E_{\mathrm{pulse}}$ is the optical pulse energy and $A$ is the $1/e^2$ beam area on the sample,
\begin{align*}
    A = \frac{\pi w_0^2}{\cos\theta}.
\end{align*}
Here, $w_0$ is the $1/e^2$ intensity radius of the circular beam in the plane perpendicular to the propagation direction,
and $\theta = \qty{70}{\degree}$ is the incidence angle measured from the surface normal. The pulse energy was measured before the beam entered the ultrahigh-vacuum chamber.

\begin{table}[htbp]
\caption{ULEEM measurement settings for the main-text and Extended Data figures. CW denotes continuous-wave operation, and n.a. denotes not applicable.}
\label{tab:measurement-settings}
\centering
\small
\setlength{\tabcolsep}{3pt}
\begin{tabularx}{\linewidth}{@{}l>{\raggedright\arraybackslash}p{0.13\linewidth}>{\raggedright\arraybackslash}Xrrr@{}}
\toprule
Data set & Figure & Description & \shortstack{$\tau_L$\\(\si{\pico\second})} &
\shortstack{$F$\\(\si{\milli\joule\per\square\centi\metre})} &
\shortstack{$E_0$\\(\si{\electronvolt})} \\
\midrule
139D03 & 2 & Static crack and focus series & 5.03 & n.a. & 6.0 \\
156D02 & 3b,c & Overview and spatially averaged transient & 6.40 & 16.20 & 10.8 \\
164D04 & 3c,d & Dynamic image sequence & 6.40 & 13.63 & 10.0 \\
164D06 & 4a inset & Long-delay edge-wave dynamics & 6.40 & 9.19 & 10.3 \\
165D01 & 4a,e and 5 & Supported-region and buried-response dynamics & 6.40 & 9.32 & 10.3 \\
150D05 & 4b,f & Locally detached edge & 5.03 & 13.62 & 15.1 \\
\addlinespace
076D01 & Extended Data 1c,e & CW dark-field image and spatial-resolution fit & CW & n.a. & 3.86 \\
150D11 & Extended Data 1b,d & Pulsed diffraction pattern and spot profiles & 5.03 & 4.84 & 57.9 \\
150D20 & Extended Data 1f & Ultrafast diffraction response & 0.25 & 8.82 & 57.9 \\
\bottomrule
\end{tabularx}
\end{table}

\FloatBarrier
\subsection{Derivation of the transport-of-intensity equation for LEEM}
To estimate the coherent image intensity near focus, we expand it to first order in defocus about the in-focus plane,
\begin{align}
I(\mathbf r,\Delta z)&\approx I(\mathbf r,0)+\Delta z\left.\frac{\partial I(\mathbf r,z)}{\partial z}\right|_{z=0}.
\end{align}
At $z=0$, the transport-of-intensity equation\cite{teague1983a} gives
\begin{align}
\left.\frac{\partial I(\mathbf r,z)}{\partial z}\right|_{z=0}
&=-\frac{\lambda}{2\pi}\nabla_\perp\!\cdot\!\left[I(\mathbf r,0)\nabla_\perp\phi(\mathbf r,0)\right].
\end{align}
Here, $\nabla_\perp=(\partial_x,\partial_y)$ and $\nabla_\perp^2=\partial_x^2+\partial_y^2$. For a uniform reflected amplitude $A_0$, with $I(\mathbf r,0)=A_0^2$, this reduces to
\begin{align}
\left.\frac{\partial I(\mathbf r,z)}{\partial z}\right|_{z=0}
&=-\frac{\lambda A_0^2}{2\pi}\nabla_\perp^2\phi(\mathbf r).
\end{align}
Substituting the height-dependent phase $\phi(\mathbf r)=4\pi h(\mathbf r)/\lambda_0$ yields
\begin{align}
  \label{eq:contrastTIE-methods}
I(\mathbf r,\Delta z) &\propto1\,-2\Delta z \frac{\lambda}{\lambda_0}\nabla_\perp^2h(\mathbf r).
\end{align}

\subsection{Image-formation model parameters}
The following calibrations and parameters were used in the forward model described in Methods.
\subsubsection*{Defocus calibration}
Following the procedure of Tromp et al.\cite{tromp2013b,tromp2013c}, we inserted a \qty{5}{\micro\m} microbeam aperture in the diagonal plane of the sample prism to form a small illumination spot on a Si(100) surface. When defocused, beams diffracted by the surface structure formed displaced images of this spot. These images were recorded at several defocus settings. The calibrated real-space image scale and known diffraction angles yielded an image-side defocus $\Delta z$ of \qty{44.6(64)}{\micro\m} per 0.001 focus-control unit.

\subsubsection*{Aberration coefficients}
We described the energy dependence of the spherical and chromatic aberration coefficients using the parameterization\cite{tromp2012b}
\begin{equation}
    C(E,E_0)=a\left(\frac{E}{E_0}\right)^b+c,
\end{equation}
where $E=\qty{15}{\kilo\electronvolt}$ and $E_0$ is the landing energy. For $C_3$, the coefficients were $a=\qty{0.024(5)}{\metre}$, $b=\num{0.27(2)}$ and $c=\qty{0.18(1)}{\metre}$. For $C_{\mathrm c}$, they were $a=\qty{-0.00174(2)}{\metre}$, $b=\num{0.490(1)}$ and $c=\qty{-0.0124(1)}{\metre}$.

\subsubsection*{Coherence parameters}
The coherence parameters were derived from the diffraction spot in Extended Data Fig.~1d. Its mean measured width, $\overline{\Delta k}_{\mathrm{FWHM}}=\qty{0.08}{\per\angstrom}$, corresponds to a Gaussian-equivalent root-mean-square (RMS) width $\sigma_k=\overline{\Delta k}_{\mathrm{FWHM}}/[2\sqrt{2\ln 2}]=\qty{340}{\per\micro\metre}$. With the column wavelength $\lambda=\qty{10.0}{\pico\metre}$ and wave number $k_0=2\pi/\lambda$, the RMS angular spread is
\begin{align*}
    \sigma_\alpha = \frac{\sigma_k}{k_0} = \frac{\sigma_k\lambda}{2\pi} = \qty{0.54}{\milli\radian}.
\end{align*}
Because the measured spot width includes instrumental broadening and depends on the fitted line shape, we used the rounded upper-bound value $\sigma_\alpha=\qty{0.60}{\milli\radian}$ in the simulations. The width along $k_y$ corresponds to an RMS energy width $\sigma_E\approx\qty{0.5}{\electronvolt}$.

\subsection{Elastodynamic simulations}
The simulations used COMSOL Multiphysics 6.3 and Dispersion Calculator v3.2, as described in Methods.
\subsubsection*{Elastic parameters}
The $z$ axis is normal to the flake surface. The TaS$_2$ stiffness components are listed in Voigt notation in the order  $(C_{11},C_{12},C_{13},C_{14},C_{33},C_{44})$. The diagonal thermal-conductivity and thermal-expansion tensors are specified by their components in the order $(11,22,33)$. Silicon is treated as isotropic. Because experimental data are limited, the elastic constants were taken from theoretical calculations. Thermal conductivities and thermal-expansion coefficients were obtained from the literature. Tables~\ref{tab:tas2-material-parameters} and \ref{tab:silicon-material-parameters} list the material parameters and their sources. To display the deformation, the simulation output was interpolated onto a $5\times5$ nm grid, with displacements amplified by a factor of 100. 

\begin{table}[htbp]
\caption{TaS$_2$ material parameters used in the simulations.}
\label{tab:tas2-material-parameters}
\centering
\small
\setlength{\tabcolsep}{4pt}
\renewcommand{\arraystretch}{1.15}
\begin{tabularx}{\linewidth}{@{}>{\raggedright\arraybackslash}p{0.31\linewidth}>{\raggedright\arraybackslash}p{0.12\linewidth}>{\raggedright\arraybackslash}X@{}}
\toprule
Quantity & Symbol & Value \\
\midrule
Independent stiffness components & $\mathbf C$ & $(178,30,11,-5,39,10), \,\si{\giga\Pa}$\cite{bastos2019} \\ %
Mass density & $\rho$ & \qty{6860}{\kilogram\per\cubic\metre} \cite{Lide2004}\\ %
Specific heat capacity & $C_p$ & \qty{313}{\joule\per\kilogram\per\kelvin} \cite{kratochvilova2017}\\  %
Thermal conductivity components & $\boldsymbol{\kappa}$ & $(5.1,5.1,1.7),\,\si{\watt\per\metre\per\kelvin}$ \cite{nunez-regueiro1985,danz2021}\\ %
Thermal-expansion components & $\boldsymbol{\alpha}$ & $(12,12,11)\times10^{-6}\,\si{\per\kelvin}$  \cite{givens1977}\\ %

\bottomrule
\end{tabularx}
\end{table}

\begin{table}[htbp]
\caption{Silicon material parameters used in the simulations.}
\label{tab:silicon-material-parameters}
\centering
\small
\setlength{\tabcolsep}{4pt}
\renewcommand{\arraystretch}{1.15}
\begin{tabularx}{\linewidth}{@{}>{\raggedright\arraybackslash}p{0.31\linewidth}>{\raggedright\arraybackslash}p{0.12\linewidth}>{\raggedright\arraybackslash}X@{}}
\toprule
Quantity & Symbol & Value \\
\midrule
Young's modulus & $E$ & \qty{170}{\giga\pascal}\cite{Hopcroft2010} \\ %
Poisson's ratio & $\nu$ & 0.28 \cite{Hopcroft2010}\\ %
Mass density & $\rho$ & \qty{2328}{\kilogram\per\cubic\metre} \cite{Hopcroft2010,Lide2004}\\ %
Specific heat capacity & $C_p$ & \qty{702}{\joule\per\kilogram\per\kelvin} \cite{Lide2004}\\ %
Thermal conductivity & $\kappa$ & $124\,\si{\watt\per\metre\per\kelvin}$\cite{Lide2004} \\ %
Thermal-expansion coefficient & $\alpha$ & $2.49 \times 10^{-6}\,\si{\per\kelvin}$ \cite{Lide2004} \\ %
\bottomrule
\end{tabularx}
\end{table}

\FloatBarrier
\subsubsection*{Dispersion Calculation}

Dispersion curves were calculated with the elastic properties and mass densities listed in Tables~\ref{tab:tas2-material-parameters} and \ref{tab:silicon-material-parameters}. The free-standing model comprised a \qty{90}{\nano\metre}-thick TaS$_2$ plate, whereas the supported model comprised a \qty{70}{\nano\metre}-thick TaS$_2$ layer perfectly bonded to a finite, \qty{10}{\micro\metre}-thick isotropic Si slab. The calculator allows modes to be classified as A-like or S-like according to their symmetry with respect to the midplane of the TaS$_2$ layer.

\subsubsection*{Optical excitation and numerical settings}

Optical penetration depths reported for 1T-TaS$_2$ vary between studies, with values of \qty{30}{\nano\metre}\cite{han2015} and \qty{45}{\nano\metre}\cite{gao2022} used at \qty{800}{\nano\metre}, and \qty{45}{\nano\metre} at \qty{1300}{\nano\metre}\cite{dean2011}. For our experimental geometry of p-polarized \qty{1030}{\nano\metre} excitation at \qty{70}{\degree} from the surface normal, the dielectric tensor determined by spectroscopic ellipsometry\cite{tiwari2026} yields an absorption depth of \qty{23.5}{\nano\metre} and a reflectivity of 0.23.

The relative populations of the excited acoustic modes depend on the absorption depth, whereas their dispersion relations are insensitive to this parameter. Given the heterogeneous specimen geometry, including partial illumination of side surfaces, we used an empirical effective absorption depth of approximately \qty{50}{\nano\metre} in the simulations, yielding good agreement with the measured relative mode intensities.

The simulated image contrast also depends on the electron-optical settings. Based on the calibrated defocus series, we used a defocus of \qty{-50}{\micro\metre}. Good agreement with the measurements was obtained for an incident fluence of \qty{8.91}{\milli\joule\per\square\centi\metre} for the supported flake (165D01), within 5\% of the calibrated experimental value (Table~\ref{tab:measurement-settings}). For the free-standing flake (150D05), we used \qty{2.61}{\milli\joule\per\square\centi\metre}. The estimated experimental fluence for this measurement was considerably higher, a discrepancy likely resulting from pump-laser misalignment.

Table~\ref{tab:simulation-geometries} lists the finite-element geometries. All external mechanical boundaries were traction-free, all thermal boundaries were insulating, and the TaS$_2$-Si interface was fully bonded. Mesh sizes were varied to balance numerical accuracy and computational cost. The final mesh spacing was \qty{2}{\nano\metre} throughout the regions analysed in Figs.~4 and 5, with a coarser mesh elsewhere, including within the Si substrate. The time-dependent solver used a relative tolerance of $10^{-3}$.

\begin{table}[htbp]
\caption{Finite-element geometries in the $x$-$z$ plane used for the simulations in Figs.~4 and 5.}
\label{tab:simulation-geometries}
\centering
\small
\setlength{\tabcolsep}{5pt}
\renewcommand{\arraystretch}{1.15}
\begin{tabularx}{\linewidth}{@{}>{\raggedright\arraybackslash}X>{\raggedright\arraybackslash}X>{\raggedright\arraybackslash}X@{}}
\toprule
Fig.~4a,c,e (supported) & Fig.~4b,d,f (free-standing) & Fig.~5b (defect) \\
\midrule
TaS$_2$ flake, \qty{10}{\micro\metre} long and \qty{70}{\nano\metre} thick, on a \qty{12}{\micro\metre}-wide and \qty{10}{\micro\metre}-deep Si substrate. The substrate extends \qty{2}{\micro\metre} beyond the flake edge. 
& TaS$_2$ flake, \qty{10}{\micro\metre} long and \qty{90}{\nano\metre} thick. 
& Same supported geometry as in Fig.~4a,c,e, with a \qty{30}{\nano\metre}-wide and \qty{2}{\nano\metre}-high rectangular cavity at the underside of the flake. \\
\bottomrule
\end{tabularx}
\end{table}
\clearpage

\clearpage
\bibliography{sn-bibliography}